\documentclass[twocolumn]{biophys-new}
\usepackage[utf8]{inputenc}
\usepackage{graphicx}
\usepackage{floatrow}
\usepackage{sidecap}

\usepackage[colorlinks,allcolors=cyan!70!black]{hyperref}

\usepackage{lipsum}
\usepackage{xr}
\title{Two timescales control the creation of large protein aggregates in cells}
\runningtitle{Biophysical Journal Template} 

\author[1,2]{A. Movilla Miangolarra}
\author[1,2]{A. Duperray-Susini}
\author[1]{M. Coppey}
\author[1,*]{M. Castellana}
\runningauthor{Miangolarra \textit{et al.}} 

\affil[1]{Laboratoire PhysicoChimie Curie, Institut Curie, PSL Research University, CNRS UMR 168, Paris, France; Sorbonne Universit\'es, UPMC Univ. Paris 06, Paris, France}
\affil[2]{These authors contributed equally}

\corrauthor[*]{michele.castellana@curie.fr}

 \papertype{Letters}

\begin{document}

\begin{frontmatter}

\begin{abstract}
Protein aggregation is of particular interest due to its connection with many diseases and disorders. Many factors can alter the dynamics and result of this process, one of them being the diffusivity of the monomers and aggregates in the system. Here, we study experimentally and theoretically an aggregation process in cells, and we identify two distinct physical timescales  that set the number and size of aggregates. The first timescale involves fast aggregation of small clusters freely diffusing in the cytoplasm, while, in the second one, the aggregates are larger than the pore size of the cytoplasm and thus barely diffuse, and the aggregation process is slowed down. However, the process is not entirely halted, potentially reflecting a myriad of active but random forces forces that stir the aggregates. Such slow timescale is essential to account for the experimental results of the aggregation process. These results could also have implications in other processes of spatial organization in cell biology, such as phase-separated droplets.
\end{abstract}

\begin{sigstatement}
Protein aggregation is a physico-chemical process that underlies many diseases and disorders, such as Alzheimer's or Huntington's disease. Here, we study experimental and theoretically the effect of a sharp decrease of diffusivity in the aggregation dynamics, such as the one that could happen in the cell due to the presence of obstacles. We find that two different timescales are important in setting the size of large aggregates and we give an estimate of the size of the aggregate at which this dramatic change in behaviour occurs, which could not be exclusive of protein aggregation but affect many other intracellular processes.
\end{sigstatement}
\end{frontmatter}

\section*{Introduction}

Protein aggregation is a  process which spans multiple order of magnitudes both in space and time: From nucleation, when a couple of monomers of a given chemical species of interest bind together to initiate the process, to the formation of large clusters containing up to millions of monomers each \cite{ravnik_aggregation_biopharma}. Among the assembly processes that are common in nature, protein aggregation in cells is of particular interest because of its role in a variety of diseases and disorders, such as Alzheimer's and Huntington's disease \cite{Selkoe2004}, and tumors
\cite{yoo2016arf6}. 
In these diseases, large micron-sized aggregates appear in cells, but it remains unclear whether toxicity is due to the intermediate-sized aggregates, or to the largest ones \cite{Ross2004}, thus indicating the utility of models and predictions for the size distribution and number of aggregates.

Here, we study theoretically and experimentally an irreversible,  diffusion-limited aggregation process using an optogenetic protein \cite{CRY2}. Irreversibility stems from a negligible fragmentation or dissolution rate of the aggregates, at least in the timescale of the experiments \cite{Taslimi2014}. The role of diffusion is also crucial, because it sets the speed at which the aggregation process unfolds: the aggregation processes in the cellular cytoplasm are influenced by the presence of physical obstacles which alter the diffusion dynamics of the aggregates within the cell \cite{Hofling_an_diff,PNAS_an_diff}.

For such irreversible aggregation processes, the only possible steady-state is the one where all proteins form a single cluster. However, this is rarely the case in biological cells, as the cytoplasm  typically exhibits multiple protein or enzymes clusters scattered all over its volume \cite{an2008reversible}. Therefore, the physical mechanisms which set the cluster number and size still remain a subject of investigation \cite{an2010dynamic,Castellana2014,buchner2013clustering}. In what follows, we address the problem of irreversible aggregation processes in human retinal pigmented epithelium cells, both theoretically and experimentally. 

 We found that two timescales control aggregation processes: One related to fast diffusion of small clusters, and another one, slower, potentially related to the hindered diffusion due to the presence of intracellular obstacles.
In this regard, in Ref. \cite{Etoc2018} it was shown that, for quasi-spherical nanoparticles in HeLa cells, the diffusivity drops by two or three orders of magnitude as the diameter of the nanoparticle is increased from $50$ to $75 \, \rm nm$, due to steric interactions with the cytosolic meshwork of the cell. Particles above this threshold---sometimes referred to as the pore size of the cytoplasm---experience almost no diffusion \cite{Luby-Phelps1987}. 
However, on longer timescales a diffusive-like movement can be observed, 
which  was related to fluctuations that stem from the incoherent effect of a network of active forces in the cell \cite{stoch_forces_cell}, such as the rearrangement of the cytoskeleton and endomembranes. Therefore large aggregates, with a radius comparable or larger than this threshold, can be thought of as being strongly confined, and subject to a dramatic hindrance in the diffusivity which may have an important effect on the aggregation dynamics.
As a result, the slow timescale above is likely to be determined by active fluctuations that affect the dynamics of intracellular objects larger than the typical pore size of the cytoplasm.



By matching theoretical predictions  with experimental results, we  estimated the threshold between these two timescales. 
Overall, our results shed light into the interplay between aggregation processes, and the dynamics of the crowded environment in the cell cytoplasm.
 
\begin{figure}[t]
\includegraphics[width=0.95\textwidth]{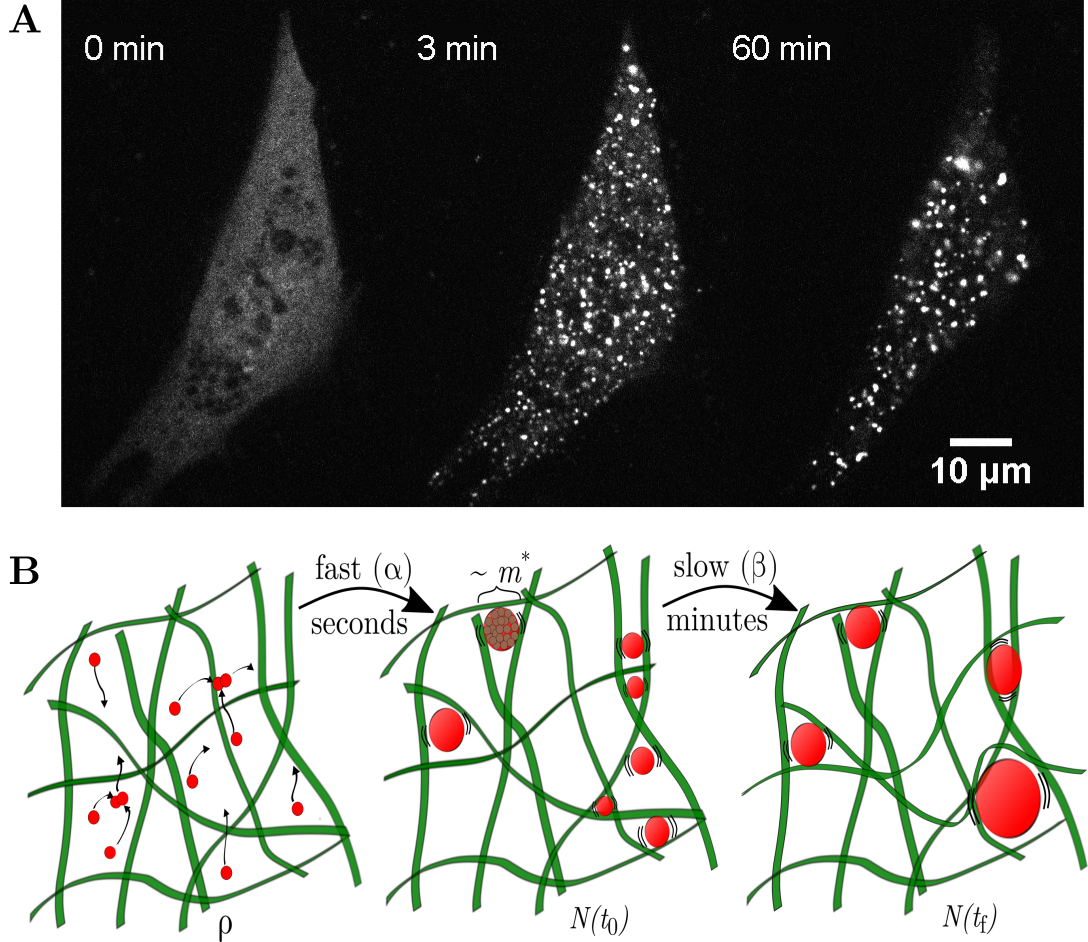}
\caption{ 
\small{Experimental system studied and theoretical model. (a) Fluorescent images of a cell expressing CRY2olig and activated with blue light every $2 \, \rm min$. The cell shows many small clusters by $t=3\, \rm min$, which then mature over time. (b) Cartoon depicting the main ingredients of the model. Left: a density $\rho$ of monomers that can freely diffuse and aggregate fast. Middle: $N(t_0)$ larger clusters of size  $\sim m^*$ or larger. Right: $N(t_{\rm f})$ large clusters are only able to move and further aggregate  stirred by active forces .}}
\label{system_fig}
\end{figure}

\section*{Experiments}

The experimental system under study is the optogenetic protein CRY2olig, which oligomerizes  upon blue light \cite{CRY2}, fused to the fluorescent tag mCherry, which is transfected into RPE1 cells ---retina pigmented epithelium 1, mammalian cells. An important feature of this optogenetic protein is the persistence of its oligomerized state even in the dark, with a half life of around $23 \, \rm mins$  in the absence of blue light exposure \cite{Taslimi2014}. 




Twenty-four hours after transfection, cells are exposed to blue light: such blue-light exposure can be regarded as an out-of-equilibrium process, which triggers protein oligomerisation. The period at which cells are exposed to blue light is $120\, \rm s$, which is significantly lower than the half life of the oligomerized state in the absence of light stimulation (23 mins), allowing us to consider the aggregation process as irreversible. The dynamics of these protein clusters are then followed for one hour with spinning disc confocal microscopy as shown in Fig. \ref{system_fig}A: From the images we extracted initial protein concentrations, final concentration of clusters and their size,  see Supplementary Material (SM) Section \ref{exp} and \ref{Image}.

The number of monomers cannot be determined directly from these images because there is a constant relating arbitrary intensity units to the monomer concentration in each pixel. An estimate of this constant was obtained by imaging  droplets with known concentration of mCherry, the fluorescent tag used in the experiments, and comparing with the imaged cells, see SM Section \ref{exp}.

Qualitatively, we were able to distinguish two different dynamical regimes. The first one is a regime characterized by rapid diffusion and aggregation, see Fig. \ref{system_fig}A, which takes place right after the blue light is switched on, and lasts for a time lapse on the order of minutes which is short compared to the imaging time of $1 \, \rm hr$. The second regime is characterized by larger clusters which exhibit slower diffusion or almost no diffusion, resulting in a slower aggregation process. 
These features are summarized in Fig. \ref{system_fig}B.
 
\begin{figure*}[t]
    \centering
   \includegraphics[width=\textwidth]{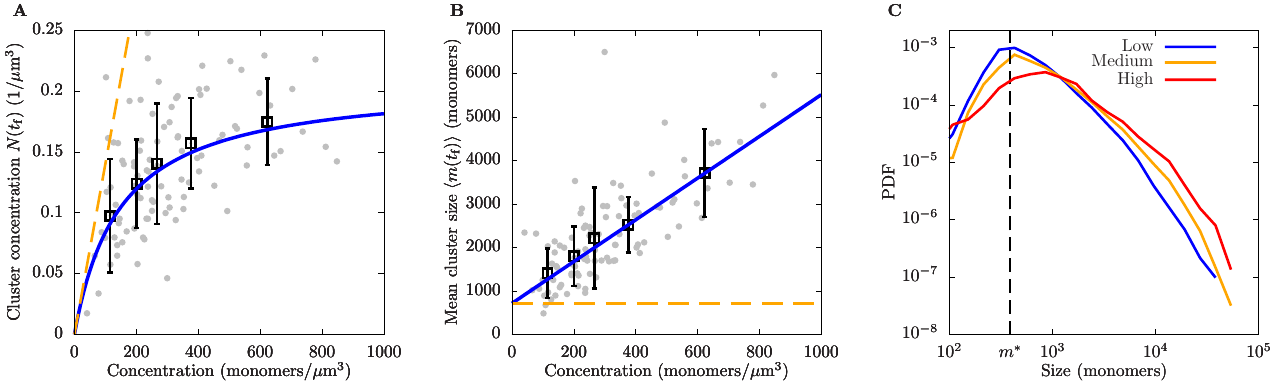}
    \caption{ \small{Results obtained for different cells $1 \, \rm hr$ after the beginning of the aggregation process. In A) and B) we plot the cluster concentration and mean cluster size as functions of the protein density $\rho$ measured in each cell, respectively. Grey dots correspond to results for individual cells, black squares to the average  over $20$ cells, and error bars to standard deviations. Blue lines correspond to least-square fits of the theoretical expressions for the cluster concentration and cluster size, that is, Eqs.  (\ref{result}) and (\ref{mean_cl}). The orange dashed lines correspond to the predictions for a passive cytoskeleton, i.e., in the absence of an active dynamics ($\beta=0$). C) Cluster-size probability density function (PDF) for different protein concentrations: Low (below $300 \, \rm  monomers/\mu m^3$), medium (between $300$ and $600\, \rm monomers/\mu \textrm{m}^3$) and high (above $600\, \rm  monomers /\mu \textrm{m}^3$). The black dashed line corresponds to the estimate of the 
cluster-mass threshold between timescales, $m^*$, obtained from a) and b).}}
            \label{main_fig}
\end{figure*}

The analysis of these images, see SM Section \ref{Image}, allowed us to obtain the cluster-size distribution, the cluster concentration,  and the mean cluster  size as functions of the initial protein concentration, see Fig. \ref{main_fig}.

\section*{Model}

The theoretical basis of irreversible aggregation processes was introduced by von Smoluchowski over a century ago, and it can be summarized into his well-known equation \cite{Smoluchowski}:

\begin{equation}
\label{Smol}
\frac{d c_i(t)}{dt}=\frac{1}{2}\sum_{j+k=i} k_{j,k} c_j(t)c_k(t) - c_i(t) \sum_k k_{i,k} c_k(t),
\end{equation} 
where $c_i(t)$ refers to the intracellular concentration of clusters with $i$ monomers, and $k_{i,j}$ is the aggregation rate between two clusters of mass $i$ and $j$, according to the law of mass action. Upon an appropriate choice of the aggregation kernel  $k_{i,j}$, Eq. (\ref{Smol}) adequately describes diffusion-limited aggregation processes. 

However, the kernel  typically does not take account of the effect of obstacles or pores, such as the ones found in the cytoplasm of a cell. 
To take account of this effect with a minimal model,  we leverage the insights from the experiments to build a kernel based on the separation of the two timescales involved in the aggregation process: On the one hand, there is a fast aggregation timescale (characterized by a rate $\alpha$), involving monomers and small clusters that diffuse rapidly, and which ultimately leads to the formation of larger agglomerates. On the other hand, there is a slow aggregation timescale (with characteristic rate $\beta$) that comprises aggregates larger than the pore size of the cytoplasm. The threshold between these two timescales is the time, $t_0$, beyond which the clusters are larger than the pore size of the cytoplasm, and we denote by $m^*$ the cluster mass at which the agglomerate attains the size of the pore of the cytoplasm and barely diffuses, see Fig. \ref{system_fig}. 

The objective of our model is to examine the effect of a sharp drop in diffusivity with particle size and, therefore, we neglect other  hydrodynamic effects, such as size-dependent diffusivity. These assumptions are supported by the findings in Ref. \cite{Etoc2018} where it was found that the most of the drop in diffusivity with particle size in cells takes place in a narrow window of size and other variations in diffusivity are small in comparison. In addition, this allows to keep the complexity of the model low, while capturing the essence of the dynamics.

\subsection*{Fast-aggregation timescale}

For the fast-aggregation timescale, we choose the following kernel in the Smoluchowski coagulation equation:
\begin{equation}
\label{const_kern}
k_{j,k}=\alpha \left [\theta (m^*-j)+\theta(m^*-k)\right],
\end{equation}
where $\alpha$ is the fast aggregation rate constant, which is assumed to be much larger than the rate constant associated with the slow aggregation timescale, and $\theta(x)$ is the Heaviside step function. Since clusters with mass larger than $m^*$ do not diffuse, they do not contribute to the aggregation rate (\ref{const_kern}). In addition, we assume that, by the time all clusters are of mass $m^*$ or larger, the contribution of the  slow process to the clustering dynamics is negligible, therefore decoupling the timescales involved in the problem.

For the kernel (\ref{const_kern}), the following change of variables is known to simplify the Smoluchowski coagulation equation \cite{Leyvraz_scaling}:
\begin{equation}
\label{change}
\varphi_i(t)=c_i(t)/N(t), \qquad d\tau=N(t) dt 
\end{equation}
where $N(t)=\sum_i c_i(t)$ and $\varphi_i(t)$ stands for the fraction of clusters that is of size $i$, which yields the following form for the Smoluchowski coagulation equations:
\begin{equation}
\label{fast_eq}
\frac{d \varphi_i(\tau)}{d \tau}=\alpha \Bigg[\sum_{j+k=i} \varphi_j(\tau)\varphi_k(\tau) \theta (m^*-j) - \varphi_i(\tau) \theta (m^*-i)\Bigg].
\end{equation}
An important feature of Eq. (\ref{fast_eq}) is its recursive structure, i.e.,  the equation for $\varphi_i$ only depends on $\varphi_j$ for 
$j < i$. One can prove inductively, see SM Section \ref{Proof}, that the solution for the mobile clusters ($\varphi_i$ for $i \leq m^*$) is given by:
\begin{equation}
\label{induction_res}
\varphi_i(\tau)=\sum_{k=1}^i (-1)^{k-1} e^{-k \alpha \tau} {i-1 \choose k-1},
\end{equation}
where we have assumed that only monomers are present at $t=0$, i.e., $\varphi_i(0)=\delta_{i,1}$. 

By summing Eq. (\ref{Smol}) for $i>0$ and using Eqs. (\ref{const_kern}, \ref{change} and \ref{induction_res}), 
we obtain a solution for the concentration of clusters, see SM Section \ref{Proof}, which reads:
\begin{equation}
\label{sol_num}
{N}(\tau)={N}(0) \exp \left \lbrace \left[ \sum_{k=1}^{m^*} \frac{(-1)^{k-1}}{k} e^{-k \upsilon \alpha} { m^* \choose k} \right]_0^\tau \right \rbrace,
\end{equation}
where the brackets denote the difference between their argument evaluated at $\upsilon = \tau$, and at $\upsilon =0$. In the limit $\tau \rightarrow \infty$ (which is equivalent to the concentration of clusters after the fast aggregation timescale has finished) and the large $m^*$ limit, the result simplifies to:
\begin{equation}
\label{Euler-Mas}
N(\tau \to \infty) \overset{m^* \rightarrow \infty}{=} N(0)\frac{e^{-\gamma}}{m^*},
\end{equation}
where $\gamma$ is the Euler-Mascheroni constant. 
It is worth noticing that in the large-time limit the results are independent of the fast aggregation constant $\alpha$: This implies that, for the  model to be consistent, we only need that the timescale separation $\alpha \gg \beta$ is satisfied.

By using Eq. (\ref{Euler-Mas}), we obtain the fraction of clusters left in the system after the fast aggregation process has finished (i.e. for $t=t_0$)
\begin{equation}
\label{exp_theor_rel}
\tilde{N}(t_0)=\frac{N(t_0)}{N(0)}=\frac{e^{-\gamma}}{m^*}.
\end{equation}
In Eq. (\ref{exp_theor_rel}) we have assumed that $m^*$ is large, which is justified by the fact that we expect $m^*$ to be of the order of $10^2$ or $10^3$, and even for $m^*=10^2$ the error in making this approximation is less than 1\%. We will now make use of Eq. (\ref{exp_theor_rel}) as the initial condition of the slow timescale. 

\subsection*{Slow-aggregation timescale}

Minutes after the start of the aggregation process, the fast aggregation process is finished (at $t \sim t_0$). Based on the experiments, we assumed that $t_0$ is small compared to the final time of the experiment $t_{\rm f}=1 \, \rm hr$, which allows us to neglect it and assume that the slow-aggregation timescale lasts for $1 \, \rm hr$ (and not $t_{\rm f}- t_0$). 

To describe the slow-aggregation regime $t>t_0$, we write a Smoluchowski coagulation equation with $k_{i,j}=\beta$, where $\beta$ is the slow-aggregation rate. The solution in this case was first given by Smoluchowski \cite{Smoluchowski}. By summing Eq. (\ref{Smol}) over all cluster sizes, see SM Section \ref{app_slow}, we obtain
\begin{equation}\label{eq_N}
\frac{dN(t)}{d t}=-\frac{\beta}{2}N(t)^2.
\end{equation}
We integrate Eq. (\ref{eq_N}) from $t=0$ to $t_{\rm f}$, substitute the density dependency in the initial conditions $N(0)=\rho \tilde{N}(t_0)$, where $\rho$ is the initial density of monomers, and obtain
\begin{equation}
\label{result}
N(t_{\rm f})=\frac{\rho \tilde{N}(t_0)}{\rho \tilde{N}(t_0) t_{\rm f} \beta/2 +1 }.
\end{equation}

Equation (\ref{result}) has two unknown parameters: $\beta$ and $\tilde{N}(t_0)$ which can be estimated from the experimental data. In particular, $\tilde{N}(t_0)$ can be obtained from relation (\ref{exp_theor_rel}).  Furthermore, using the relationship $\rho=N(t_{\rm f})\langle m(t_{\rm f})\rangle$, one can estimate  the mean cluster mass as a function of the protein density:
\begin{equation}
\label{mean_cl}
\langle m(t_{\rm f})\rangle=\rho\, t_{\rm f} \beta/2 + \tilde{N}(t_0)^{-1}.
\end{equation}
Equations (\ref{exp_theor_rel}) , (\ref{result}) and (\ref{mean_cl}) constitute our main theoretical results. 

\subsection*{Comparison with the experimental data}

In order to test the model, we fit Eqs. (\ref{result}) and (\ref{mean_cl}) to the experimental data for the mean cluster mass  and cluster density at the end of the experiment. Results are shown in Fig. \ref{main_fig}A and B.
The fit  yields  $\beta=9.6 \,\textrm{hr}^{-1}\, \mu \textrm{m}^3$ and $\tilde{N}(t_0)=1.4 \times 10^{-3}$. Using Eq. (\ref{exp_theor_rel}), one obtains a value of $m^*=390$ monomers for the mass threshold above which clusters are expected to be trapped in the cytoplasm.

The experimental data allowed us to quantify also the cluster-size distribution, see Fig. \ref{main_fig}C. Our estimate of $m^*$ is close to the peak of the cluster-size distribution: This result is consistent with the assumptions made in the model,  that the aggregation process is slowed down for clusters of mass above $m^*$, see Fig. \ref{main_fig}C. 

We can assess the consistency of our result with other experimental data by estimating the pore size of the cytoplasm from the prediction for $m^*$. In this regard, in the framework of diffusion-limited cluster aggregation (DLCA) \cite{Kolb_clusters1983}, it has been suggested that the fractal dimension for DLCA in the presence of restructuring  is $d_f=2.18$ \cite{Meakin_restructuring1988}, i.e., 
\begin{equation}
\label{fractal_agg}
\Bigg(\frac{R_{m^*}}{r_0}\Bigg)^{d_f} \sim  \,m^*
\end{equation} 
where $R_{m^*}$ is the radius of an aggregate of mass $m^*$ and 
$r_0 = 2.5 \, \rm nm$ is the radius of an individual CRY2olig monomer, i.e., the average size of a protein containing $\sim 500$ residues (see SM Section \ref{fractal_dim} for details). Therefore, we obtain for the radius of an aggregate of mass $m^*$: $R_{m^*}\sim 39\,\textrm{nm}$. It should be noted that these calculations are correct up to a constant that we cannot determine. Nevertheless, our estimates are consistent with the threshold found in Ref. \cite{Etoc2018}, where  the threshold between diffusing and non-diffusing particles is reported to be between $25$ and $37.5\, \rm nm$. 

\section*{Conclusion}

In this Letter, we studied diffusion-limited aggregation of an optogenetic protein, CRY2olig,  in mammalian cells, combining an experimental and a theoretical approach.


Our main result is the identification of two different timescales in the aggregation process: On the one hand, there exists a short timescale where small clusters can freely  diffuse and aggregate, leading to the formation of larger agglomerates. 
On the other hand, later on,  large aggregates barely diffuse or do so very slowly. Based on previous work \cite{Etoc2018, Luby-Phelps1987,stoch_forces_cell}, this effect could be largely due to confinement within the cytoskeleton and other cytosolic obstacles: as a result, large clusters cannot diffuse nor aggregate, unless the confining obstacles move or rearrange on a longer timescale.
The predicted threshold between the two timescales corresponds to cluster sizes of $\sim 400$ monomers, or $\sim 39\, \rm nm$ of radius, which roughly corresponds to the cytosolic pore size \cite{Etoc2018}.  

Our model yields a quantitative estimate of the aggregation rate, $\beta$,  relative to the long time scale:  This rate would characterize the incoherent dynamics of an intracellular network of active forces, such as molecular motors \cite{stoch_forces_cell}, which could thus be regarded as an active stirring of the aggregates. 

In addition, our analysis demonstrates that clustering of CRY2olig in mammalian cells is markedly different from aggregation in a  passive material with a fixed pore size, where the dynamics of the aggregation would halt as soon as aggregates reach the pore size. This comparison was made in Fig. \ref{main_fig}A and B, where the orange dashed lines represent the predictions for a passive material with the same pore size as that of the cells in our experiment ($\beta = 0$), while solid blue lines represent our model prediction, which includes the active stirring of clusters. 

The ideas developed in this study can be generalized to a variety of biological systems that reach a steady state driven by out-of-equilibrium processes, such as synthesis, degradation, traffic or recycling of proteins \cite{Turner2005}. In addition, the mechanisms identified here could be extended to the kinetics of other intracellular phenomena, such as liquid-liquid phase separation \cite{Gueroui2019}. Indeed, systems under binodal phase separation might exhibit as well two time scales in their coarsening dynamics.
The fast timescale rate, $\alpha$, would represent the diffusion-limited coalescence of droplets in the early kinetics. On the other hand, as droplets grow and diffusion slows down, the main driving force of coarsening would presumably be Ostwald ripening, whose details could be taken into account by a parameter, or function, equivalent to the slow aggregation rate, $\beta$. 
Given that there is a free-energetic cost for a droplet to deform around a network of obstacles \cite{Shin2018}, the effect of obstacles in diffusion would become important only for droplets with a characteristic radius $R_{m^*}$ or larger. Thus, we expect the values of $R_{m^*}$ and $m^*$ to be similar to the ones predicted by our analysis.

\section*{Author Contributions}

\footnotesize A. M. M. and M. Castellana, developed the theoretical model. A. M. M. performed the image analysis. A. D.-S. and M. Coppey designed and performed the biological experiments. M. Castellana and M. Coppey conceived the study.

\section*{Acknowledgments}

This study is supported PIC3i grant from Institut Curie and by Agence nationale de la recherche  (ANR) grant ANR-17-CE11-0004. 
We acknowledge M.-H. Stern for his contribution to the PIC3i project regarding oncogenic signaling, and for valuable conversations. 
We also acknowledge 
R. Botet, F. Brochard, J. Prost, J.-F. Joanny, T. Risler, S. Bell and P. Sens for valuable discussions. 
The authors thank the Imaging Nikon Center (PICT-LM) and A. El Majou, head of the Recombinant Proteins Platform, at Institut Curie. M. Coppey thanks the Labex CelTisPhyBio (ANR-10-LBX-0038) and Idex Paris Sciences et Lettres (ANR-10-IDEX-0001-02 PSL), the France-BioImaging infrastructure supported by ANR Grant ANR-10-INSB-04 (Investments for the Future), and Institut Pierre-Gilles de Gennes (Laboratoire d’excellence, Investissements d’Avenir program ANR-10-IDEX-0001-02 PSL and ANR-10-LABX-31).

\bibliography{bibliography}


\onecolumn

\section*{Supplemental Material}
\vspace{1cm}

\section{Experimental Procedures}\label{exp}

\subsection{Cell culture}
The immortalized hTERT RPE1 cells (Human Retinal Pigmented Epithelium) were cultivated in DMEM F12 without Phenol Red (Gibco, Life Technologies) supplemented with 10\% Fetal Bovine Serum (FBS) without antibiotic, hereafter called the growth medium. They were maintained at 37$^{\circ}$C in humidified atmosphere with 5\% $\textrm{CO}_2$, tested and certified as mycoplasma free.

\subsection{Transitory cells transfection by Cry2Olig-mCherry}
RPE1 cells were detached by trypsin and centrifuged for $3\, \rm min$, $100\, \rm g$ at room temperature to eliminate it. The pellet was kept and resuspended on  growth medium. They were transfected on suspension by jetPrime (Polyplus transfection), with $1\, \rm \mu g$ of DNA plasmid vector Cry2Olig-mCherry (purchased from Addgene, number 60032), and then platted on fluorodishes. According to the recommendation of manufactory, the medium was replaced after four hours by a fresh one. From there, the manipulation of cells was done in the complete dark.

\subsection{Quantitative estimation of fluorescent protein concentration}
To estimate the concentration of proteins in cells using the fluorescent signal, we calibrated the intensity on the camera using mCh-6His protein purified at $4.19\, \rm mg/ml$ (a gift from El Marjou. A, Platform of Curie Institute). We performed serial dilutions of the stock solution (1, 1:2, 1:4, 1:8, 1:10, 1:16, 1:32, 1:64, 1:100, 1:128, 1:1000) in the cell growth medium, and the medium alone was used for background estimation. For each dilution, we put a drop of $10\, \rm \mu l$  into a fluorodish and we imaged the drop using the exact same parameters as for the cell imaging experiments. Two images were acquired at a focus right above the coverslip, as for cell imaging. We then quantified the average fluorescent intensity using Fiji. The total intensity of the image was background subtracted and averaged over the size of the whole image. Data were plotted and gave rise to a linear relationship between raw intensities of the images and concentrations of recombinant fluorescent proteins.  We fitted data with a line and used the value of the slope to convert intensities into concentrations.     

\subsection{Optogenetic experiments }
All experiments were performed using 100x objectives (oil immersion, numerical aperture 1.4) by Inverted Spinning Disk Confocal Roper/Nikon, EMCCD 512x512 evolve (pixel size: $16\, \rm \mu m$) photometrics come from to Imaging Nikon Center (PICT-LM) in Curie Institute. Live imaging was on normal growth condition and preserved by Life Imaging Service Yokogawa head: CSU-X1 integrated in Metamorph software by Gataca Systems. 
Twenty-four hours after transfection, cells were kept at 37$^{\circ}$C and were imaged before any activation with blue light over 17 z-stack ($0.5\, \rm \mu m$) at $561\, \rm nm$ ($0.134\, \rm mW$). The same cells were imaged at the end of the activation routine using the same 17 z-stacks while keeping the same focus. Optogenetic activations were performed every two minutes for a total duration of one hour, using the laser blue light at $491 \, \rm nm$ ($0.506\, \rm mW$). We selected cells for further image quantification based on their visible viability, on their presence in the field of view at the end of the experiment (some cells escaped the field of view after one hour), and on the absence of pixels saturation (very bright, saturated clusters could appear over the time course of the experiment). All laser settings and parameters of the camera (time of exposition, gain) were kept constant for all experiments and calibration of the concentration.

\section{Image Analysis} \label{Image}

The initial concentration of the protein is obtained from the cell image at the initial time, $t=0$. The cell is separated from the background and the intensity is computed as the average of the intensity in the cell after subtracting the background intensity, using Matlab \cite{matlab2018}. We estimated the volume of the cells by measuring the area of the cell just above the coverslide and assuming an effective height such that the total intensity of the 3D final image equals the total intensity of this 2D initial image times this effective height. This effective height parameter varies from cell to cell and has a mean value of 1.1$\mu$m and a standard deviation of 0.4$\mu$m. 

In  order to quantify the size and frequency of the cluster at $t=1 \, \rm hr$, we smoothed the image with a gaussian filter, substracted the mean background intensity, located the local maxima of intensity in the image, and performed a watershed transform to estimate the spatial extent of each cluster \cite{watershed}. The size of the clusters is then determined by considering that the cluster is composed of the pixels that have at least one fifth of the intensity of the maximum of such cluster. 
In addition, we considered a bright spot to be a cluster only if the intensity of its peak is at least 2000 arbitrary units above the background intensity---which corresponds to peaks with at least  $\sim 20$ monomers. 
Once the clusters are located and their boundaries defined, we add up the total intensity of each of them, separately, to obtain an estimate of the mass of each cluster, i.e., the total number of monomers in each of the clusters. This number might be slightly underestimated due to a potential self-quenching effect of the fluorescent tag upon aggregation.

\section{Parameter fitting} \label{Fits}

The two datasets that we want to fit with Eqs. (\ref{result}) and (\ref{mean_cl}), i.e., cluster density and mean cluster mass, have different units and numerical values. In what follows, we will introduce a least-square minimization such that, when minimising the squares to find the best fitting parameters, both datasets are equally taken into account. To achieve this, we introduce  
\begin{equation}\label{xx}
  \sum_i \left\{ \frac{1}{\mu_1}[f_1(x_i) - y_i^{(1)}]\right\}^2 + \sum_i \left\{ \frac{1}{\mu_2} [f_2(x_i) - y_i^{(2)}]\right\}^2, 
\end{equation}
where $f_{1,2}$ are defined by Eqs. (\ref{result}) and (\ref{mean_cl}), the 2-tuples $(x_i,y_i)^{(1,2)}$ denote each of the datapoints $i$ of each dataset (1 or 2, cluster concentration or cluster size), and $\mu_{1,2}$ are  the mean values of the datapoints of each dataset: $\mu_{1,2}={M_{1,2}}^{-1} \sum_{i=1}^{M_{1,2}} y_i^{(1,2)}$, $M_{1,2}$ being the number of datapoints.

\section{Solution for the fast-aggregation timescale} \label{Proof}

Given that Eq. (\ref{fast_eq})  is a recursive equation for $\varphi_k$, in what follows we will attempt an inductive proof of the solution for any $\varphi_k$ for $k \leq m^*$, for which the Heaviside step function is equal to one. 

In what follows, we will show that, for $i\leq m^*$, if the ansatz (\ref{induction_res}) holds for  $\varphi_1, \cdots, \varphi_{i-1}$, then it holds for $\varphi_i$ as well. To achieve this, we insert the ansatz (\ref{induction_res}) in Eq. (\ref{fast_eq}), where we evaluate
\begin{align}
  \sum_{j=1}^{i-1} \varphi_j(\tau) \varphi_{i-j}(\tau)&= \nonumber \\
  \sum_{j=1}^{i-1} \sum_{n_1=1}^j  \sum_{n_2=1}^{i-j} (-1)^{n_1+n_2-2}
  e^{-(n_1+n_2) \tau \alpha} { j-1 \choose n_1-1} { i-j-1 \choose n_2-1}&= \nonumber \\ 
  \sum_{j=1}^{i-1} \sum_{s=2}^i    \sum_{n_2=1}^{s-1} (-1)^{s-2}
  e^{-s \tau \alpha} { j-1 \choose s-n_2-1} { i-j-1 \choose n_2-1}&,
\end{align}
where, in the last equality, we have made the change of variable $s=n_1+n_2$. Now we can apply Vandermonde's identity
\begin{equation}
  {m+n \choose r}=\sum_{k=0}^r {m \choose k}{n \choose r-k},
\end{equation} 
which yields
\begin{align}\nonumber
  \sum_{j=1}^{i-1} \varphi_j(\tau) \varphi_{i-j}(\tau)=& \sum_{j=1}^{i-1} \sum_{s=2}^{i}  (-1)^{s-2}  e^{-s\tau \alpha} {i-2 \choose s-2}\\
  =& \nonumber (i-1) \sum_{s=2}^{i}  (-1)^{s-2}  e^{-s\tau \alpha} {i-2 \choose s-2}.
\end{align} 
Equation (\ref{fast_eq}) for $\varphi_i, \quad i\leq m^*$ now reads, assuming the ansatz (\ref{induction_res}) for $\varphi_j, \quad j<i$,
\begin{equation}\label{temp}
  \frac{d \varphi_i(\tau)}{d \tau}= \alpha (i-1) \sum_{s=2}^{i} (-1)^{s-2}  e^{-\alpha s\tau} {i-2 \choose s-2} - \alpha \varphi_i(\tau) 
\end{equation}
which can be rewritten as follows
\begin{equation}
\frac{d (\varphi_i(\tau) e^{\alpha \tau})}{d \tau}= \alpha (i-1) \sum_{s=2}^{i} (-1)^{s-2}  e^{-\alpha (s-1)\tau} {i-2 \choose s-2}
\end{equation}
and solved by direct integration along with the monodisperse initial conditions (which make the constant from the integration vanish), 
yielding
\begin{equation}
  \label{ind_res}
  \varphi_i(\tau)=\sum_{s=1}^i (-1)^{s-1} e^{-s \alpha \tau} {i-1 \choose s-1}.
\end{equation}
Alternatively, Eq. (\ref{ind_res}) can be recast into the form:
\begin{equation}
  \varphi_i(\tau)=\frac{(1 - e^{-\alpha \tau})^i}{ e^{\alpha \tau}-1}
\end{equation}
by the binomial theorem. 

The number of clusters as a function of our rescaled time $\tau$ can be obtained summing over all $i$ Eq. (\ref{Smol}) with the kernel (\ref{const_kern}):
\begin{align}
  \sum_i \frac{dc_i(t)}{dt}=&-\alpha \frac{1}{2}\sum_{i,j+k=i} c_j(t)c_k(t)\left [\theta (m^*-j)+\theta(m^*-k)\right] - \alpha  \sum_{i,k} c_i(t) c_k(t) \left [\theta (m^*-i)+\theta(m^*-k)\right] \nonumber \\ =&-\alpha \sum_{i,j+k=i} c_j(t)c_k(t) \theta (m^*-j)\nonumber -\sum_i\Bigg [ \alpha c_i(t) N(t) \theta (m^*-i)+\alpha c_i(t)\sum_{k=1}^{m^*}c_k(t)\Bigg ] \nonumber \\=&-\alpha N(t) \sum_{j=1}^{m^*} c_j(t).
\end{align}
Using Eq. (\ref{change}) in the main text, we obtain the equation for the cluster concentration in the rescaled time $\tau$
\begin{equation}
  \label{number_fast}
  \frac{d N(\tau)}{d \tau}=-\alpha N(\tau) \sum_{i=1}^{m^*} \varphi_i(\tau). 
\end{equation}

Using the so-called \textit{hockey-stick identity}:
\begin{equation}
  {m^* \choose n}=\sum_{k=0}^{m^*-n} {k + n -1 \choose n-1},
\end{equation}
after inserting Eq. (\ref{ind_res}) into Eq. (\ref{number_fast}) we obtain
\begin{align}
  \frac{d {N}(\tau)}{d \tau}=& - \alpha N(\tau) \sum_{i=1}^{m^*} \sum_{j=1}^{i} (-1)^{j-1} e^{-j \alpha \tau} {i-1 \choose j -1} \nonumber \\=& - \alpha N(\tau) \sum_{j=1}^{m^*} \sum_{i=0}^{m^*-j} (-1)^{j-1} e^{-j \alpha \tau} {i+j-1 \choose j -1} \nonumber \\=&- \alpha N(\tau) \sum_{j=1}^{m^*} (-1)^{j-1} e^{-j \alpha \tau} {m^* \choose j},
\end{align}
whose solution is Eq. (\ref{sol_num}) in the main text.
Finally, we give some more details regarding the appearance of the Euler-Mascheroni constant in Eq. (\ref{Euler-Mas}) of the main text. We start from Eq. (\ref{sol_num}) in the main text:
\begin{equation}
{N}(\tau)={N}(0) \exp \left \lbrace \left[ \sum_{k=1}^{m^*} \frac{(-1)^{k-1}}{k} e^{-k \upsilon \alpha} { m^* \choose k} \right]_0^\tau \right \rbrace,
\end{equation}
which, in the limit $\tau \rightarrow \infty$ takes the form
\begin{equation}
  N(\tau \to \infty) =  N(0) \exp \left \lbrace -\sum_{k=1}^{m^*} \frac{(-1)^{k-1}}{k} { m^* \choose k} \right \rbrace,
\end{equation}
which can be rewritten as
\begin{equation}
  \label{infty_lim}
  N(\tau \to \infty) = \frac{ N(0)}{m^*} \exp \left \lbrace -\sum_{k=1}^{m^*} \frac{(-1)^{k-1}}{k} { m^* \choose k} + \log m^* \right \rbrace.
\end{equation}
We can identify
\begin{equation}
  \sum_{k=1}^{m^*} \frac{(-1)^{k-1}}{k} { m^* \choose k}
\end{equation}
as the $m^*$-th harmonic number, $H_{m^*}$ which diverge logarithmically as $m^* \rightarrow \infty$. The Euler-Mascheroni constant is the difference between the $m^*$-th harmonic number and the logarithm of $m^*$ in the limit where  $m^* \rightarrow \infty$:
\begin{equation}
  \gamma= \lim_{n\rightarrow \infty} \left[ \sum_{k=1}^{n} \frac{(-1)^{k-1}}{k} { n \choose k} - \log n \right].
\end{equation}  Hence, Eq. (\ref{infty_lim}) can be recast as
\begin{equation}
  N(\tau \to \infty) = \frac{ N(0)}{m^*} e^{-\gamma},
\end{equation}
which is Eq. (\ref{Euler-Mas}) in the main text, where $\gamma$ is the Euler-Mascheroni constant.

\section{Solution for the slow-aggregation timescale}
\label{app_slow}
Starting from the Smoluchowski equation, with the kernel $k_{j,k}=\beta$:
\begin{equation}
  \frac{d c_i(t)}{dt}=\frac{\beta}{2}\sum_{j+k=i} c_j(t)c_k(t) - \beta c_i(t) \sum_k c_k(t),
\end{equation}
we sum over all $i$ and rewrite the constraint on the summation $j+k=i$ with a Kronecker delta, yielding
\begin{align}
  \sum_i \frac{d c_i(t)}{dt}=&\frac{\beta}{2}\sum_{i,j,k} \delta_{j+k,i}c_j(t)c_k(t) - \sum_i \beta c_i(t) \sum_k c_k(t) \nonumber \\
  =&\frac{\beta}{2}\sum_{j,k} c_j(t)c_k(t) - \sum_i \beta c_i(t) \sum_k c_k(t),
\end{align}
and with the definition $N(t)=\sum_i c_i(t)$ we obtain Eq. (11) of the main text:
\begin{equation}
  \frac{dN(t)}{d t}=-\frac{\beta}{2}N(t)^2.
\end{equation}
The solution to this equation is 
\begin{equation}
  N(t)=\frac{1}{t+C}
\end{equation}
where, imposing the initial condition $N(0)=\rho \tilde{N}(t_0)$, $C$ takes the value $(\rho \tilde{N}(t_0))^{-1}$, yielding Eq. (\ref{result}).

\section{Estimate of the sizes of the aggregates}\label{fractal_dim}
In this section we give an estimate of the characteristic size of the aggregates based on the Diffusion-Limited Cluster-Cluster Aggregation (DLCA) framework. This theory assumes that clusters diffuse freely and bind to each other as soon as they come to contact \cite{Kolb_clusters1983}. If the bonds created by each binding event are rigid and maintain their shape the resulting structure will be very sparse. However, in many cases this may not be true and  the bonds may rearrange to make a more compact structure. 
Taking into account this rearrangement, the fractal dimension of the DLCA clusters can be taken to be $d_f=2.18$ \cite{Meakin_restructuring1988}. Combining this result with Eq. (\ref{fractal_agg}), we obtain a characteristic radius of the aggregate. This radius stands for the typical size of the aggregates, and it does not strictly represent the radius of an aggregate, nor implies that the aggregate has spherical shape.

\end{document}